\documentclass{PoS}

\title{Astronuclear Physics with Short-Lived Isotopes}

\ShortTitle{Astronuclear Physics with Short-Lived Isotopes}

\author{\speaker{C.A. Bertulani}\thanks{carlos.bertulani@tamuc.edu}\\
Department of Physics and Astronomy, Texas A \& M University - Commerce, Commerce, TX 75429, USA\\
        E-mail: \email{carlos.bertulani@tamuc.edu}}

\abstract{A revolution in  nuclear physics is underway. If you know hadron physics you also know that it will  last long, as most past developments in nuclear physics have shown. It will take many decades of dedicated efforts of theorists and experimentalists to give astronomers, cosmologists and stellar modelers what they need for the accuracy required in the description and modeling of many cosmic phenomena of interest. In this contribution I review  a small number of examples of the utmost relevance for the {\it nuclear in the astro} and in particular the humongous importance of the new area of astronuclear physics studied in radioactive beam facilities.}

\FullConference{X Latin American Symposium on Nuclear Physics and Applications (X LASNPA),\\
		1-6 December 2013\\
		Montevideo, Uruguay}
		
\begin{document}
 
\section{Nuclei in the Cosmos}
\subsection{ Why do stars shine and why do they live so long?} 
Astronomy is one the oldest of all sciences. Early catalogue, identification of position, motion and brightness of planets and stars have long been used for practical applications such as a measure of time, distance, and navigation. But only in the last hundred years a thorough understating of stars was possible with the advent of nuclear physics. 

Legend says that Hans Bethe, somewhen during the 1930's was admiring the stars with his girlfriend when he mentioned that he was  the only person on Earth (hopefully not in the Universe) that understood how stars shine. I can only imagine that his girlfriend  was disappointed by his lack of a more romantic, mundane, and {\it poetic view of nature}. But Bethe was right and the immense light intensity and energy released by stars are powered by nuclear reactions at very low energies, despite the incredibly high temperatures in stellar cores. Bethe knew that there is a very large energy gain by fusion of 4 protons into the $^4$He nucleus ($\alpha$-particle). In the process, two of the protons change into two neutrons by the emission of positrons (beta-decay) and  28 MeV ($\sim 4.5 \times 10^{-12}$ J) is gained for each $\alpha$-particle produced. Bethe also proposed that nuclei of carbon and nitrogen were catalyzers of the $\alpha$-particle, eating 4 protons along several steps of a chain reaction, spitting out the $\alpha$-particle in the end. The  catalyzers are recycled  back to the same state as they were before, much like enzymes catalyze chemical reactions in biological processes. Bethe's proposal \cite{Bet39} is now known as the {\it CNO cycle} (O stands for oxygen) . This process is indeed the driving mechanism for energy generation in massive, hot stars. 
Stars such as our sun catalyze $\alpha$s by a sequence of proton-induced fusion reactions  known as the {\it pp-chain} \cite{Ad11}. 

Stars only exist due to the most subtle phenomenon predicted by quantum mechanics: the {\it tunnel effect}. The first application of this prediction was in nuclear physics by a study  of the lifetime of nuclei decaying by $\alpha$ emission. Some nuclei, such as $^{210}$Po emit $\alpha$s by tunneling through the Coulomb barrier, a process lacking explanation until George Gamow used the tunneling concept to explain and calculate $\alpha$-emission half-lives \cite{Gam28,GC28}. Without the tunneling effect the stars would burn much too quickly and {\it the Universe as we know would not exist}.  

\subsection{Why there is enough carbon for life?}  
Here is another classic example of the need of a nuclear physics course for astronomy majors. This helps them understand basic facts about stars and  prepare them to do a good job in astrophysics research. Nuclear physics made a breakthrough in understanding the {\it origins of life} with the discovery of the triple-alpha process by which three alpha particles react to form the carbon nucleus. This was a purely theoretical hypothesis based on the understanding of how nuclei interact. The triple-alpha process is known as a pathway to heavier elements but the non-resonant triple-alpha reaction is  too small due to the very improbable collision of three $\alpha$-particles in a stellar environment. Sir Fred Hoyle proposed that the observation of plenty of carbon in the Universe, responsible for the existence of life on Earth, must imply a resonance in the continuum of berillium (not bound) and in carbon close its alpha emission threshold, more precisely at 7.65 MeV  \cite{Hoy54}. Following this prediction, William Fowler and collaborators performed an experiment at the Kellogg Radiation Laboratory confirming the existence of the resonance, or {\it Hoyle state}. This state is known as the {\it nuclear  life state} and some think of it as a support to the {\it anthropic principle}: we exist, therefore carbon must allow this state \cite{Fow84,Hjo11}. The triple-alpha reaction continues to attract genuine interest from the nuclear physics community, both experimentally \cite{Fyn05} and theoretically \cite{Hjo11}.

\subsection{Why do I need to know about low energy nuclear scattering?} 
The dawn of nuclear physics was crucial for the development of many areas of science, besides cosmology and stellar physics.  In 1989, I recall attending a talk in Germany in which a speaker mentioned the nucleon-nucleon {\it scattering length}. The audience, mostly nuclear physicists, seemed to have forgotten this concept. ``How come you don't know what is a scattering length?" -  the speaker complained. ``This was invented by a nuclear physicist (Hans Bethe again, in 1949 \cite{Bet49})!" - he added. At the time, the nuclear physics community was mostly interested to prove the theoretical prediction of a phase transition from nucleons to  a quark-gluon plasma occurring in central collisions of {\it relativistic heavy ions} \cite{PBM07}. Such a phase transition is thought to occur in the {\it core of neutron stars}. This project required so much effort and dedicated research on quantum chromodynamics (QCD) of quarks and gluons system, that few nuclear physicists had time or interest on low energy nuclear physics. 

Since the development of the effective range expansion (EFE) by Hans Bethe only a few formal developments using the ERE was done up to three decades ago, most notably by another nuclear physicist, Herman Feshbach at MIT. His seminal work on closed and open channels applied to the so-called {\it Feshbach resonances} is a standard tool to understand new phenomena associated with few-body physics and the physics of few atoms trapped in a cavity \cite{Fes58}. Nowadays, scattering length and {\it effective range} are part of the everyday jargons used  in atomic physics and quantum optics. They are basic knowledge for the foundations of quantum phenomena, {\it optical lattices}, {\it Bose-Einstein condensates}, or the {\it unitary Fermi gas},.

\subsection{Where are the quarks and gluons in nuclear physics?}  If nuclear physics is so important for  stars, why aren't there more breaking news in nuclear physics? The answer is simple. Nuclear physics is very difficult and is, arguably,  {\it the most difficult problem in all physics}. The crucial issues are: (a) nucleons are composite objects, (b) nucleon-nucleon interactions are residuals of the strong interactions of quarks and gluons inside the nucleons - it is not known at the level of the Coulomb interaction so fundamental for chemistry and condensed matter physics - and (c) the nuclear many-body problem involves only a few (up to $\sim 200$) nucleons disallowing simplifications so common in other areas of physics. It is very difficult for nuclear physicists to obtain results within the level of accuracy sometimes required by many astrophysics phenomena.   Nuclear interactions are also strongly influenced by the presence of other nucleons within a nucleus, or extended nuclear matter as in the interior of a neutron star. 

With the advent of fast computers, new methods have been developed to improve our knowledge of nuclear forces. Based on many developments in particle and nuclear physics in the last decades, we are pretty certain that QCD is the underlying theory describing the interactions between all hadrons. Predicting nuclear properties by solving  the nuclear many-body Schr\"odingier equation has proven itself very difficult during the last 80 years. Hence, the task of describing nuclear properties starting from the QCD Lagrangian will require a computational effort beyond our present resources.   This is true despite the humongous progress made by solving QCD on a lattice  during the last three decades \cite{Du08}.  Predictions based on the extrapolation of present computing resources leads us to estimate that it might take another 80 years to obtain nuclear properties strictly by using  QCD.

Maybe the full QCD machinery is not necessary to calculate most of the reactions of astrophysical interest. Some restricted information from lattice QCD would be more than enough to describe low energy nuclear physics. This is the basic idea of {\it effective field theories} (EFT) \cite{Wei79}.  The low energy constants derived from QCD can be used either to separate energy scales in nuclei participating in a reaction, or by renormalizing the nucleon-nucleon interaction in the nuclear medium.  Applications of these ideas to key nuclear reactions of interest for solar physics and for other stellar evolution scenarios have been obtained recently  \cite{Rup00,BHK02,Rup11,Rup13}. For example,  calculations of radiative capture reactions using EFT and lattice gauge theory  Refs. \cite{Rup11,Rup13} and a calculation for the triple-alpha reaction based on the {\it chiral-EFT}  \cite{Epe13}.

\section{New nuclei  in the chart}

\subsection{Halo nuclei and Efimov states.} A simple but ingenious experiment reported in 1985 by Isao Tanihata and his group on interaction cross sections of light nuclei close to the drip line was the seed of a new era in nuclear physics \cite{Tan85}. This experiment and others following it, have shown that some nuclei such as $^{11}$Li possess a long tail neutron distribution. Evidently, the long tail is due to the low binding energy of the valence nucleons. Typically, neutron binding energies are of the order of 8 MeV for most stable nuclei. But neutron separation energies as low as a few keV have been observed in short-lived nuclei. The long (visible in a logarithmic scale) tail of the loosely bound neutrons constitute a phenomenon now known as "halo" and the nuclei with this property are known as {\it halo nuclei}. If that is all, why is it so interesting? Because of correlations. For example, $^{10}$Li and two neutrons do not form a bound system, but put them together in $^{11}$Li and they will bind by a little ($\sim 300 $ keV). $^{11}$Li got so much attention that it was difficult in the 1990's to pass a week without a preprint on some sort of calculation or experiment involving it. Physicists working with three-body systems saw in this nucleus much of what had been predicted decades before on the delicate structures arising from loosely bound three-body systems. For them, $^{11}$Li was (is) the prototype of an effect known as the {\it Efimov effect}, the appearance of bound states in a system composed of unbound two-body subsystems  \cite{Efi70,Zhu93}. Nuclear physicists called them {\it Borromean systems}, in connection with the heraldic symbol of the aristocratic Borromeo family from the fifteen century \cite{Zhu93}. Borromean rings are composed of three intertconnected rings, such that if one is cut loose, the remaining two also get free. 
A large activity in the area of three and few-body physics have profited from this early work on nuclear physics in the 1980's and 1990's.  Efimov states in borromean systems  have become a common feature in atomic \cite{Kra06} and nuclear \cite{BH07} physics.

\subsection{Magic numbers.} In 1949 Maria Goeppert-Mayer and Hans  Jensen  proposed that nucleon forces should include  a spin-orbit interaction able to reproduce the nuclear magic numbers 2, 8, 20, 28, 50, 82, 126, clearly visible from the systematic study of nuclear masses.  Now we know that as nuclei move away from the line of stability ($Z\sim N$, where $Z$ ($N$) is the proton (neutron) number) those magic numbers might change due to correlations and details of the nucleon-nucleon interaction, e.g., the {\it tensor-force}  \cite{Ste13,Ots05}. These  and other properties of nuclei far from the stability and in particular those close to the {\it drip-line} (i.e., when adding one more nucleon leads to an unbound nucleus) increased enormously the interest for low-energy nuclear physics, in particular for the  prospects of its application to other areas of science \cite{Cot10}. The applications in astrophysics are evident. For example, the {\it rapid neutron capture process} (rp-process) involves nuclei far from the stability valley, many of which are poorly, or completely, unknown. 

It is not easy to predict what one can do with short-lived nuclei on earth. Sometimes, they have lifetimes of milliseconds, or less. It is tricky to explain to the laymen the importance of such discoveries in the recent history of nuclear physics. For example, in 1994, the unstable doubly-magic nucleus $^{100}$Sn was discovered at the Gesellschaft für Schwerionenforschung (GSI), Darmstadt, in Germany \cite{Sch94}. At the same time the laboratory developed a new {\it cancer therapy} facility based on the stopping of high energy protons in cancer cells. To celebrate the new facility, the laboratory invited high-ranked authorities from the German government. They surely brought along the main news media in the country. Wisely, the laboratory management included the discovery of $^{100}$Sn alongside the proton therapy facility in the celebration agenda. At one moment a reporter asked a physicist what is $^{100}$Sn good for practical purposes, what was followed by a long silence and a curious answer: "I think one can develop better, lighter, cans." Maybe so, but the can would last about 1 second! The only sure prediction for applications of short-lived nuclei is that it is impossible to understand well stellar evolution without a dedicated study of their properties.

\subsection{New nuclear physics accelerators.} With so much of science at stake it was not difficult to obtain funding for new nuclear physics laboratories with the sole goal of studying nuclei far from the stability line. At the beginning of 1990's the world's most active intermediate energy ($\sim 100-1000$ MeV/nucleon) nuclear physics laboratories at GSI (Germany), GANIL (France), RIKEN (Japan) and MSU (USA) changed gears to produce secondary beams of unstable nuclei through fragmentation reactions. The new era in nuclear physics started to gain momentum \cite{BCH93}.  It is worthwhile mentioning that before it, most physics studied in those facilities had to do with central collisions  with the purpose to study the {\it equation of state} (EOS) of nuclear matter at high densities and temperatures \cite{Li08}. This EOS (pressure versus density), specially at low temperatures, is crucial for  understanding the physics of supernova explosions and neutron stars. In fact, masses and radii of neutron stars are constrained by the EOS of nuclear matter.  In contrast to the study of EOS,  the physics of radioactive secondary beams was mostly driven by peripheral, direct reactions. Stripping, Coulomb and nuclear excitation, two- and three-body breakup, and other direct reactions were and remain the principal tools to access spectroscopic information of interest to model and develop theories for short-lived nuclei, far from the stability line.  Their relevance for astrophysics purposes is discussed next.

\section{Astrophysics with secondary beams of unstable nuclei}

\subsection{Very small cross sections}
The large temperatures in stellar cores translate to small nuclear kinetic energies, compared to typical accelerator energies. For example, nuclear reactions in the sun's core occur at relative energies of around 10 keV. To measure the cross sections at such low energies requires special kinds of nuclear accelerators. There exists in fact techniques to accelerate (or decelerate) unstable nuclei to such small energies, but not many facilities are able to do it. And since the cross sections are very small at the relevant energies, high intensity beams are required, a very difficult task specially with short-lived nuclei. 

\subsection{Inverse kinematics}
Reactions at inverse kinematics use a stable target (e.g., a proton gas target) to access information on the desired cross sections  or other nuclear properties of interest at the low astrophysics energies. In order to extract this information one has to rely heavily on reaction theory. A few of these theories and their challenges are listed next.

\subsubsection{Coulomb excitation and breakup} 

{\bf Pigmy resonances.} Below the Coulomb barrier (a few MeV/nucleon) nuclei interact mostly via the Coulomb force, which is well known. Above the Coulomb barrier there are differences between the Coulomb and the nuclear interaction contributions to the cross sections which allows them to be disentangled  \cite{BB88}. The Coulomb part can be well described theoretically if first-order perturbation theory is enough. Many nuclear spectroscopic information of relevance for astrophysics have been obtained in this way. For example, one of the subjects of relevance for nuclear astrophysics is the excitation of low energy resonances above the particle emission threshold, the so-called {\it pigmy resonances} \cite{SAZ13}. One believes that such excitations can absorb a non-negligible amount of energy during {\it cataclysmic processes}, e.g., during a supernovae explosion, which can lead to a dramatic change in the outcome of the {\it abundances of the heavy elements}  \cite{Go98}.

{\bf Radiative capture reactions.} When a projectile is loosely-bound and is dissociated into two parts ($a \rightarrow b+c$) in the Coulomb field of a large-$Z$ target, an analysis of the Coulomb breakup allows for the inference of the inverse process $b+c \rightarrow a$ of relevance for nuclear astrophysics, e.g., for {\it radiative capture} cross sections at low energies. A large number of radiative capture cross sections have been investigated in this way, e.g., the $^7$Be(p,$\gamma$)$^8$B cross section and reaction rate of relevance for high energy production in the solar core. 

The so-called {\it Coulomb dissociation method} is now a successful indirect tool to obtain information on radiative capture reactions of astrophysical interest \cite{BBH86}. The detailed balance theorem allows a direct relation between the dissociation process and the capture cross section. But many reactions involving loosely-bound nuclei require the consideration of multiple step processes. For example, during the breakup the loosely-bound projectile can make excursions in the continuum and iterate with the Coulomb field of the target  many times before the fragments get asymptotically free. Multistep processes in the continuum led to the development of new techniques, such as the {\it continuum-discretized coupled-channels} (CDCC) method, e.g., as applied to a loosely-bound nucleus in Ref. \cite{BC93}. The challenge is to develop reliable nuclear models to obtain the scattering states including resonances. 

{\bf Microscopic methods.} The {\it resonating group method} (RGM) is perhaps the most advanced tool to describe nuclear scattering and fusion of light nuclei at low energies. It was originally introduced by Hill and Wheeler in 1953 \cite{HW53}. It has been the tool of choice to calculate cross sections for reactions involving cluster-like nuclei such as $^{12}$C($\alpha$,$\gamma$)$^{16}$O \cite{LK85}.  To stress its relevance for cosmology, if the cross section for this reaction is, e.g., twice the value one assumes for it,  a 25 solar masses star will not produce $^{20}$Ne since carbon burning ceases. Such an oxygen rich star is more likely to collapse into a black hole while carbon rich progenitor stars is more likely to leave behind a neutron star \cite{WW93}.  The RGM has been combined with elaborated structure models such as the {\it no-core shell model} (NSCM) to obtain good estimates of nuclear fusion reactions.  An example was the recent calculation of the $^3$H(d,n) cross section to explain the experimental data obtained at the National Ignition Facility \cite{NQ12}.
There is a large effort to combine microscopic methods of low energy astrophysical reactions with the Coulomb dissociation method analysis based on the CDCC method to obtain a better description of reactions studied in the radioactive beam facilities \cite{NRQ11}. 

\subsection{Trojan Horse method.} 
The Trojan Horse Method (THM) is based on the idea that due to off-shell effects a nucleus $x$ can be brought to react at low energies with a heavy target and overcome the Coulomb barrier if it is carried inside another nucleus with a much larger kinetic energy \cite{Bau86}. The THM has obtained relevant information on several reactions of interest for nuclear astrophysics \cite{Spi11}. A Catania experimental group has proven that the energy dependence of the cross sections obtained with the THM is  the same as those obtained in direct methods \cite{Spi13}. Thus, a normalization of THM results to those obtained with direct methods allows the indirect measurement of new data points at energies below those known so far. Applications of the THM range from reactions of interest for the production of light elements in numerous stellar environments to reactions of interest for {\it big bang nucleosynthesis} (BBN) \cite{Piz14}. 

\subsection{Asymptotic normalization coefficients.} 
This method relies on the experimental analysis of classical nuclear reactions such as elastic scattering and peripheral transfer reactions with extrapolation of the experimental scattering phase shifts to the low astrophysical energies. Such information is embodied in the same overlap function as the amplitude of the corresponding astrophysical reaction of interest \cite{Xu94}. This is just because the reaction is peripheral, probing the tails of the bound-state wave functions. In fact, most astrophysical nuclear fusion processes reed only this information for an accurate theoretical calculation. 

The tail of the overlap functions (or wave functions) contain information on the complex many-body physics in the nuclear interior through a single parameter; the {\it asymptotic normalization coefficient} (ANC) which multiplies a well known Whittaker (for charged particles) or modified Bessel (for neutron-induced reactions) functions. The ANC method is an effective theory to obtain cross sections of reactions of interest in astrophysics, in which the ANCs
embody the information on  short distances and many-body aspects so difficult to quantify \cite{Tri14}.

\subsection{Surrogate reactions.}
Many (n,$\gamma$) reactions are of interest for the {\it rapid neutron capture} process. Many of them will never be measured directly at the astrophysical energies. As with the THM,  valuable (n,$\gamma$)  cross sections can be obtained via {\it surrogate reactions} in which the neutron is carried along by a nucleus to react with a target \cite{CB70,ED10}. Perhaps the simplest of these are (d,p) reactions, due to the simplicity of the deuteron structure.  One of the problems is the fact that the neutron in a nuclear environment carries angular (and other) quantum numbers which are not the same as free neutrons. Unless the angular momentum plays a minimal role in the reaction, this brings  difficulties in extracting the desired neutron-induced reaction from the surrogate equivalent. Theoretically, this is the same as the implication that the Hauser-Feshbach formalism for (n,$\gamma$) reactions agrees with the Ewing-Weisskopf formalism \cite{CI10}. The first include angular momentum and the second does not.

For (d,p) reactions  another challenge is the description of the reaction in terms of qualified three-body models. A proper description of such reactions  requires a correct treatment of the interaction between its constituents to all orders. There is a large effort going on to describe the mechanism with the {\it Alt-Grassberger-Sandhas equations} \cite{AGS67}, difficult to solve numerically. 

\subsection{Knockout reactions.}
Single-nucleon knockout reactions  are a tool for studying single-particle occupancies and correlation effects in the nuclear shell model. In such experiments a  fast, mass A, projectile collides  a light nuclear target yielding residues with mass (A - 1) \cite{Baz09}.  Knockout reactions have been useful to reduce the uncertainties in astrophysical reactions. For example,  the neutron removal from a radioactive ion beam can be used to populate the nuclear states of interest. A practical example is  $^{33}$Ar, where excited states were measured with uncertainties of several keV. This lead to an improvement of 2 orders of magnitude  in the uncertainty of level energies and a 3 orders of magnitude improvement in the uncertainty of the $^{32}$Cl(p,$\gamma$)$^{33}$Ar reaction rate, important for the astrophysical modeling of the {\it rapid proton capture}, or rp-process, e.g. in X-ray burst from a neutron star accreting matter from a companion
in a stellar binary system \cite{Cle04,Scha05}.

\subsection{Charge exchange reactions.} Charge exchange reactions involve reactions such as (p,n)  or ($^3$He,t) reactions.  They are often used to obtain information on weak interaction processes  which cannot be extracted from direct beta-decay experiments \cite{Sas12}. In many astrophysics phenomena such as  {\it supernovae core collapse}, temperatures and densities are high enough to ensure that electron capture in nuclei reduces the electron fraction in the medium, and drive the nuclear composition to more neutron rich and heavier nuclei. Nuclei with $N >40$ tend to dominate the matter composition for densities larger than a few $10^{10}$ gcm$^{-3}$.  In core-collapse supernovae simulations, electron capture stops at such densities and the capture becomes entirely due to free protons. 
Charge-exchange reactions help to infer the matrix elements for nuclear response by spin and isospin operators involved in electron capture processes and also in neutrino-scattering reactions occurring in stellar environments \cite{MLD00}. 

\section{What are the key reactions in nuclear astrophysics?}

This is one of the most common questions raised by those often unsatisfied with large number of reactions quoted by nuclear astrophysicists claiming to progress our understanding of stellar evolution and other cosmic objects. Can't you narrow them down, by a measurement or calculation of  two or a few crucial reactions? {\bf The answer is  NO}. I have tried to give a few examples of key reactions across this text, but {\bf stars enjoy diversity}. 

As a consequence, astronomers and astrophysicists have to patiently wait  for the painstaking progress in nuclear physics to get a firm basis for many stellar phenomena of interest. Many reactions will be studied during the next decades in billion dollar nuclear physics facilities already built and under construction  around the world. Arguably, reactions with short-lived nuclear isotopes will constitute one of the last frontiers assessed in nuclear physics experiments \cite{BG10}.

\bigskip

The author acknowledges support under U.S. DOE Grant DDE- FG02- 08ER41533.

\end{document}